\documentclass[12pt]{article}
\usepackage{graphicx}

\newsavebox{\sboxpubnumber} \newsavebox{\sboxpubdate}
\newcommand{\pubdate}[1]{\begin{lrbox}{\sboxpubdate}{#1}\end{lrbox}}

\newcommand{\Title}[1]{\begin{center} {\Large #1 } \end{center}}
\newcommand{\Author}[1]{\begin{center}{ \sc #1} \end{center}}
\newcommand{\Address}[1]{\begin{center}{ \it #1} \end{center}}

\newenvironment{Abstract}{\begin{quotation}  }{\end{quotation}}
\newenvironment{Presented}{\begin{quotation} \begin{center}
             PRESENTED AT\end{center}\bigskip
      \begin{center}\begin{large}}{\end{large}\end{center}
      \end{quotation}}
\newcommand{\Acknowledgements}{\bigskip  \bigskip \begin{center} \begin{large}
             \bf ACKNOWLEDGEMENTS \end{large}\end{center}}

\begin{document}

\newcommand{\gtrsim}{ \mathop{}_{\textstyle \sim}^{\textstyle >} }
\newcommand{\lesssim}{ \mathop{}_{\textstyle \sim}^{\textstyle <} }

\begin{titlepage}
\pubdate{\today}                    

\vfill
\Title{Moduli problem and Q-ball baryogenesis in
  gauge-mediated SUSY breaking models}
\vfill
\Author{Masahiro Kawasaki}
\Address{Research Center for the Early Universe, School of Science,
         University of Tokyo \\ 
         Hongo 7-3-1, Bunkyo-ku, Tokyo 113-0033, Japan}
\vfill

\begin{Abstract}
We investigate whether the Affleck-Dine mechanism can produce
sufficient baryon number of the universe in the gauge-mediated 
SUSY breaking models, while evading the cosmological moduli problem by
late-time entropy production. We find that the Q-ball formation
makes the scenario difficult, irrespective of the detail mechanism
of the entropy production.
\end{Abstract}
\vfill
\begin{Presented}
    COSMO-01 \\
    Rovaniemi, Finland, \\
    August 29 -- September 4, 2001
\end{Presented}
\vfill
\end{titlepage}
\def\thefootnote{\fnsymbol{footnote}}
\setcounter{footnote}{0}

\section{Introduction}
\label{sec:intro}

In superstring theories, there generally exist various dilaton and
modulus fields. These fields (we call them "moduli" ) are
expected to acquire masses of the order of the gravitino mass
$m_{3/2}$ through some non-perturbative effects of supersymmetry
(SUSY) breaking. It is well known that the moduli cause
serious cosmological problem~\cite{Coughlan} because they have only
gravitationally suppressed interactions with other particles and hence
have long lifetimes. The moduli with mass $O(100)$ GeV decay at the Big
Bang Nucleosynthesis (BBN) epoch and spoil the success of the BBN by
destroying the synthesized light elements, while moduli with lighter
mass ($\lesssim 1$~GeV) may overclose the universe or emit
X($\gamma$)-rays  giving  too many contributions to the cosmic
background radiation~\cite{Kawasaki}. 

The mass of moduli ($=$ gravitino
mass $m_{3/2}$) depends on models of SUSY breaking. In hidden sector
models~\cite{Nilles} the SUSY breaking in the hidden sector is
mediated by gravitation and SUSY particles (squarks, sleptons, etc )
in the observable sector as well as gravitino obtain the mass of weak
scale $\sim O(100)$~GeV. On the other hand, in gauge-mediated SUSY
breaking models~\cite{Giudice}, the gauge interactions mediate the
SUSY breaking effects. In gauge-mediated SUSY breaking, the gravitino
cannot acquires mass by the gauge interactions but only through
gravitation. Thus, the mass of gravitino is much lighter ($\lesssim
1$~GeV) than that in the hidden sector models.  Here we consider 
the moduli problem in the gauge-mediated SUSY breaking models.

In order to avoid the moduli problem, we need some huge entropy
production process by which the moduli density is diluted.  So far the 
most successful mechanism for entropy production is ``thermal
inflation'' proposed by Lyth and Stewart~\cite{Lyth}.  The thermal
inflation model in the gauge-mediated SUSY breaking models was
intensively investigated in Refs.~\cite{Hashiba,Asaka1,Asaka2} where
it was shown that the thermal inflation can solve the moduli problem.
However, any process that dilutes the moduli also dilutes primordial
baryon asymmetry of the universe. Thus, we must produce sufficiently
large baryon asymmetry before the entropy production occurs. 
The promising candidate for mechanism of such efficient baryon number
generation is the Affleck-Dine (AD) baryogenesis~\cite{Affleck}. In
fact, it was shown in Ref.~\cite{Asaka2} that both the present baryon
asymmetry and small moduli density can be explained by the thermal
inflation and the Affleck-Dine mechanism for the gauge-mediated SUSY
breaking models.

However, it has been found that the dynamics of the Affleck-Dine
baryogenesis is complicated by the existence of Q
balls~\cite{Kusenko,Enqvist}. In the gauge-mediated SUSY breaking
models the potential for the Affleck-Dine field becomes flat at large
amplitudes. For such a flat potential the Q-ball formation is
inevitable~\cite{Kasuya1,Kasuya2,Kasuya3}. In order to produce a large
baryon number the initial amplitude of the AD field should be large,
which also leads to the formation of Q balls with huge baryon($=$Q)
number.  Since large Q balls are stable, the baryon number may be
confined in the form of Q balls and there may exist very small baryon
asymmetry in the cosmic plasma, which means that the baryogenesis does
not work.

Unstable Q balls can provide all the charges created before the charge 
trapping by the produced Q balls, but rather small amplitudes of the
AD field are necessary for the Q balls to decay into the ordinary
baryons, nucleons. Thus, sufficient baryon number is not created from
the beginning.

Therefore we study the cosmological moduli problem and baryogenesis in
the gauge-mediated SUSY breaking models taking into account the Q-ball
formation. It is found that the Q balls seriously affect the Affleck-Dine
baryogenesis and lower its efficiency\footnote{
The detailed analysis is found in Ref.~\cite{KKT}. }

\section{Moduli Problem}

The modulus field $\eta$
obtain a mass of the order of the gravitino mass $m_{3/2}$. During the
primordial inflation the modulus field is expected to sit at some
minimum of the effective potential determined by the K\"ahler
potential and the Hubble parameter. In general, the minimum during the
inflation deviates from the true minimum of the moduli potential at
low energies and the difference of the two minimum is considered to be
of the order of the gravitational scale $M(=2.4\times 10^{18}
\mbox{GeV})$. After the inflation, when the Hubble parameter becomes
comparable to the mass of the modulus, the modulus field begins to
roll down toward the true minimum and oscillates. Then, the modulus
density is estimated as
\begin{equation}
   \label{eq:moduli-dens}
   \frac{\rho_{mod}}{s} \simeq \frac{1}{8}T_{RH}
   \left(\frac{\eta_0}{M}\right)^2,
\end{equation}
where $s$ is the entropy density, $T_{RH}$ is the reheating
temperature and $\eta_0$ is the initial amplitude of the modulus
oscillation ($\eta_0 \sim M$). In deriving Eq.(\ref{eq:moduli-dens}),
we have assumed that the modulus mass is equal to $m_{3/2}$ and the
reheating takes after the moduli field starts the oscillation. Since
$T_{RH}$ should be higher than about 10~MeV to keep the success of the
BBN, the moduli to entropy ratio is bounded from below,
\begin{equation}
   \frac{\rho_{mod}}{s}\gtrsim 1.25\times 10^{-3}{\rm GeV}.  
\end{equation}

The decay rate of the modulus is very small because it has only
gravitationally suppressed interaction. The lifetime is roughly
estimated as
\begin{equation}
    \tau_{\eta} \sim 10^{18}\sec 
    \left(\frac{m_{3/2}}{100{\rm MeV}}\right)^{-3}.
\end{equation}
Thus, for $m_{3/2} \lesssim 100$~MeV, the lifetime is longer than the
age of the universe and its present density much larger than the
critical density which is given by 
$\rho_c/s_0 = 3.6\times 10^{-9} h^2 {\rm GeV}$,
where $h$ is the present Hubble parameter in units of
100km/sec/Mpc and $s_0(\simeq 2.8\times 10^3$~cm$^{-3})$ is the
present entropy density. The modulus with larger mass (100~MeV
$\lesssim m_{3/2} \lesssim $ 1~GeV ) decays into photons whose flux
exceeds the observed background X(or $\gamma$)-rays. Therefore the
modulus is cosmological disaster and should be diluted by some entropy
production process.

\section{Affleck-Dine mechanism and Q-ball formation}

In the Minimal Supersymmetric Standard Model (MSSM), there exist flat
directions, along which there are no classical potentials. Since flat
directions consist of squarks and/or sleptons, they carry baryon
and/or lepton numbers, and can be identified as the Affleck-Dine
field. These flat directions are lifted by SUSY breaking effects. In
the gauge-mediated SUSY breaking models, the potential of a flat
direction is parabolic at the origin, and almost flat beyond the
messenger scale: 
\begin{equation}
    V_{gauge} \sim \left\{ 
      \begin{array}{ll}
          m_{\phi}^2|\Phi|^2 & \Phi \ll M_S \\
          M_F^4 \log \frac{|\Phi|^2}{M_S^2}
          & \Phi \gg M_S \\
      \end{array} \right.,
\end{equation}
where $M_{S}$ is the messenger mass scale.

Since the gravity always exists, flat directions are also lifted by
the gravity-mediated SUSY breaking effects:
\begin{equation}
    V_{grav}=m_{3/2}^2 \left[ 1+K
      \log \left(\frac{|\Phi|^2}{M} \right)\right] |\Phi|^2,
\end{equation}
where $K$ is the numerical coefficient of the one-loop
corrections. This term can be dominant only
at high energy scales because of the small gravitino mass 
$\lesssim O(1\mbox{GeV})$. 

In addition, there is a thermal effect on the potential, which appears
at two-loop order, as pointed out in Ref.~\cite{AnisimovDine}. This
comes from the fact that the running of the gauge coupling $g(T)$ is
modified by integrating out heavy particles which directly couples
with the AD field. This contribution is given by
\begin{equation}
     V_T \sim T^4 \log\frac{|\Phi|^2}{T^2}.
\end{equation}

The baryon number is usually created just after the AD field starts
coherent rotation in the potential, and its number density $n_B$ is
estimated as 
\begin{equation}
    n_B(t_{osc}) \simeq \varepsilon \omega \phi_{osc}^2,
\end{equation}
where $\varepsilon(\lesssim 1)$ is the ellipticity parameter, which
represents the strongness of the A-term, and $\omega$ and $\phi_{osc}$ 
are the angular velocity and the amplitude of the AD field at the
beginning of the oscillation (rotation) in its effective potential. 

Actually, however, the AD field feels spatial instabilities
during its coherent oscillation, and deforms into nontopological
solitons, Q balls \cite{Kusenko,Enqvist,Kasuya1}. 
From numerical calculations~\cite{Kasuya1,Kasuya3}, Q balls absorb almost
all the baryon charges which the AD field obtains, and the typical
charge is estimated as \cite{Kasuya3}
\begin{equation}
    Q \simeq \beta \left(\frac{\phi_{osc}}{M_F}\right)^4,
\end{equation}
where $\beta \approx 6 \times 10^{-4}$. Consequently, the present
baryon asymmetry should be explained by the charges which come out of
the Q balls through the evaporation, diffusion, and decay of Q
balls.

In the case of the unstable Q balls, they decay into nucleons and light
scalar particles. Since the temperature at the BBN time is very low
($\sim 1$ MeV), Q balls cannot decay into light scalars. This rate is
thus given by \cite{Coleman}
\begin{equation}
    \frac{dQ}{dt} \lesssim \frac{\omega^{3} A}{192 \pi^{2}},
\end{equation}
where $A$ is a surface area of the Q ball.

In the case of the stable Q balls, the evaporation is the only way to
extract the baryon charges from Q balls. The total evaporated charge
from the Q ball is estimated as \cite{Laine,Banerjee,Kasuya3}, 
\begin{equation}
   \label{eq:chargeevp}
   \Delta Q \sim 10^{15}
   \left(\frac{m_{\phi}}{\mbox{TeV}}\right)^{-2/3}
   \left(\frac{M_F}{10^6\mbox{GeV}}\right)^{-1/3} Q^{1/12}.
\end{equation}
Hence the baryon number density is suppressed by the factor 
$\Delta Q/Q$, in comparison with the case of no stable Q-ball
production.

On the other hand, where $V_{grav}$ dominates the potential at larger
scales, the gravity-mediation type Q balls (`new' type) are produced
\cite{Kasuya2}, if $K$ is negative, while, if $K$ is positive, it is
not until the AD field enters $V_{gauge}$ dominant region that it
feels instabilities, and the gauge-mediation type Q balls are produced
(the delayed Q balls) \cite{Kasuya3}. Notice that the sign of $K$ is
in general indefinite in the gauge-mediated SUSY breaking models.

When the AD field starts to oscillate in the $V_{grav}$-dominant
region, where $H_{osc} \sim \omega \sim m_{3/2}$, the baryon number is
produced as $n_B \simeq \varepsilon m_{3/2}\phi_{osc}$. For the
negative $K$ case, the `new-type' Q balls are created, and its charge
is written as
\begin{equation}
    \label{eq:new-Q}
    Q \simeq \tilde{\beta} \left(\frac{\phi_{osc}}{m_{3/2}}\right)^2,
\end{equation}
where $\tilde{\beta} \simeq 6\times 10^{-3}$. This type of the Q ball
is also stable against the decay into nucleons, and the amount of the
baryons in the present universe is explained by the charge evaporation
from the Q balls. The charge evaporated from the Q ball is estimated
as \cite{Kasuya2} 
\begin{equation}
    \label{eq:new-evap}
    \Delta Q \sim 2.2 \times 10^{20} 
    \left(\frac{m_{3/2}}{100 \mbox{keV}}\right)^{-1/3}
    \left(\frac{m_{\phi}}{\mbox{TeV}}\right)^{-2/3}.
\end{equation}

\section{Baryogenesis and the moduli problem}

As we mentioned in the Introduction, the late-time entropy production
necessary for the dilution of the moduli also dilutes the baryon
numbers created earlier very seriously, but the sufficient numbers
could remain, if the Q-ball production is not taken into account. We
will see that the Q-ball formation put very serious restriction on the 
efficiency of the AD baryogenesis, whether the
produced Q balls are stable or not.

\subsection{Stable Q balls}
 
Since the baryon number is supplied only by the evaporation from
stable Q balls, it should be suppressed by the factor $\Delta Q/Q$,
compared with no Q-ball formation. We will show that this fact 
considerably reduces the power of the AD baryogenesis.


\subsubsection{Gauge-mediation type Q balls when the zero-temperature 
potential is dominated}

If the $V_{gauge}$ dominates over the thermal logarithmic potential 
$V_T$, $M_F \gtrsim T_{osc}$, so that the initial amplitude is
constrained as $\phi_{osc} \gtrsim (T_{RH}/M_F)^2 M$.
$V_{gauge}$ also dominates over $V_{grav}$, which leads to the
condition   $\phi_{osc} \lesssim M_F^2/m_{3/2}$.
Combining these two equations, we have $T_{RH} \lesssim M_F^2/\sqrt{m_{3/2}M}$.

From the stability ($M_Q/Q < 1$~GeV) and survival condition ($\Delta Q < Q$), the maximal baryon-to-entropy ratio is 
\begin{eqnarray}
    Y_B & \lesssim & 6.2 \times 10^{-27} \varepsilon
    \left(\frac{\Omega_{mod}h^2}{0.2}\right)
    \left(\frac{m_{\phi}}{\mbox{TeV}}\right)^{-2/3}.
\end{eqnarray}
Therefore, the baryon-to-entropy ratio is too small to explain 
the present value of the order $10^{-10}$.


\subsubsection{Gauge-mediation type Q balls when the thermal
logarithmic potential is dominated}

In this case, $V_{T}$ is dominant over $V_{gauge}$, and the AD
field starts to oscillate when $H_{osc} \sim T_{osc}^2/\phi_{osc}$. We
should also use the charge of the formed Q ball as $Q\simeq \beta
(\phi_{osc}/T_{osc})^4$. Therefore, the fraction of the evaporated
charge is written as
\begin{eqnarray}
    \frac{\Delta Q}{Q}  & \sim & 8.9 \times 10^{17}
    \left(\frac{m_{\phi}}{\mbox{TeV}}\right)^{-2/3}
    \left(\frac{M_F}{10^6 \mbox{GeV}}\right)^{-1/3}
    \left(\frac{T_{RH}}{M}\right)^{11/3}
    \left(\frac{\phi_{osc}}{M}\right)^{-11/2}.
\end{eqnarray}
Then the baryon-to-entropy ratio becomes
\begin{eqnarray}
    Y_B & \simeq & 5.3 \times 10^{-10} \varepsilon
    \left(\frac{\Omega_{mod}h^2}{0.2}\right)
    \left(\frac{m_{\phi}}{\mbox{TeV}}\right)^{-2/3}
    \left(\frac{M_F}{10^6\mbox{GeV}}\right)^{-1/3}
    \left(\frac{T_{RH}}{M}\right)^{5/3}
    \left(\frac{\phi_{osc}}{M}\right)^{-3/2}.
\end{eqnarray}

\begin{figure}[t!]
    \centering
    \includegraphics[height=7cm]{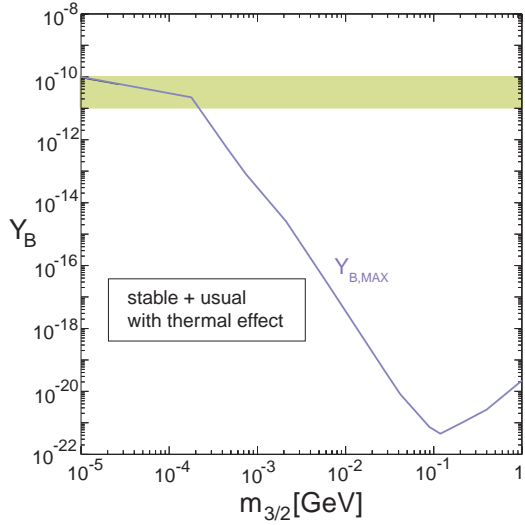}
    \caption[fig2]{\label{fig:stusu} 
    Largest possible baryon-to-entropy ratio in the stable Q-ball
    scenario in the thermal logarithmic potential.  }
\end{figure}

We plot the baryon-to-entropy ratio in the function of $m_{3/2}$ in
Fig.~\ref{fig:stusu}. We can thus marginally explain the present value
($Y_B \gtrsim 10^{-11}$) for $m_{3/2} \lesssim 100$ keV, but the
reheating temperature is $\sim 2.0 \times 10^{13}$ GeV, so high for
natural inflation models to provide.  It is noticed that $Y_B$ is 
maximized at $\Delta Q \sim Q$.

\subsubsection{Delayed Q balls when the zero-temperature 
potential is dominated}

Both the AD field and moduli fields start to oscillate when
$H_{osc}\sim m_{3/2}$, the ratio of the baryon number and the
energy density of the moduli stays constant to the present,
the baryon number is given by 
$n_B \simeq \varepsilon m_{3/2} \phi_{osc}^2 \Delta Q/Q$,
and the baryon-to-entropy ratio becomes
\begin{eqnarray}
    \label{eq:etab}
     Y_B & \sim & 2.8 \times 10^{-24} \varepsilon
     \left(\frac{\Omega_{mod}h^2}{0.2}\right)
     \left(\frac{m_{\phi}}{\mbox{TeV}}\right)^{-2/3}
     \left(\frac{m_{3/2}}{100\mbox{keV}}\right)^{8/3}
     \left(\frac{M_F}{10^6\mbox{GeV}}\right)^{-4}
     \left(\frac{\phi_{osc}}{M}\right)^2.
\end{eqnarray}

\begin{figure}[t!]
    \centering \includegraphics[height=7cm]{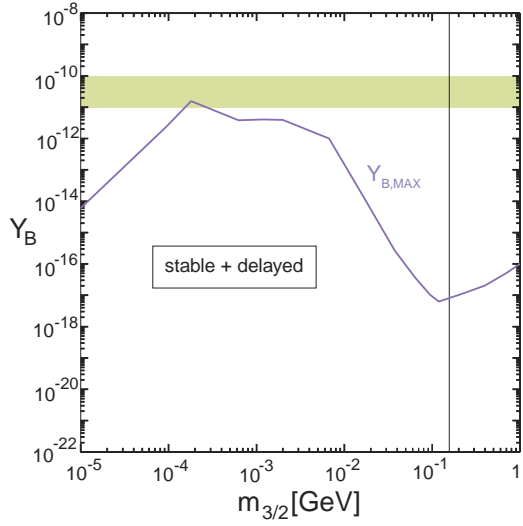}
    \caption[fig3]{\label{fig:delayed} 
    Largest possible baryon-to-entropy ratio in the stable delayed
    Q-ball scenario. Notice that it can be applied in the
    zero-temperature logarithmic potential $V_{gauge}$ for $m_{3/2}
    \lesssim 0.16$ GeV, while all ranges of $m_{3/2}$ can be applied
    in the thermal logarithmic potential $V_T$.  }
\end{figure}

Figure \ref{fig:delayed} shows the maximum value of the
baryon-to-entropy ratio. As can be seen, this scenario is marginally
successful ($Y_B \sim 10^{-11}$) only for $m_{3/2} \sim 200$
keV. Notice that, the value of $Y_B$ is the same as that in the
thermal logarithmic potential, to be considered in the next
subsection, for $m_{3/2} \lesssim 0.16$ GeV. 

\subsubsection{Delayed Q balls when the thermal
logarithmic potential is dominated}

In the case that $V_T$ is dominant over
$V_{gauge}$, the maximum value of the
baryon-to-entropy ratio is also shown in Fig.~\ref{fig:delayed}. 
We can see that $Y_B$ is marginally enough ($Y_B \sim 10^{-11}$)
only for $m_{3/2} \simeq 200$ keV. Notice that the largest possible
value of $Y_B$ is achieved when
$\Delta Q \sim Q$.


\subsubsection{New type Q balls}
From Eqs.(\ref{eq:new-Q}) and (\ref{eq:new-evap}), we have
\begin{equation}
    \frac{\Delta Q}{Q} \simeq 6.1 \times 10^{-23}
    \left(\frac{m_{\phi}}{\mbox{TeV}}\right)^{-2/3}
    \left(\frac{m_{3/2}}{100\mbox{keV}}\right)^{5/3}
    \left(\frac{\phi_{osc}}{M}\right)^{-2}.
\end{equation}
This leads to the baryon-to-entropy ratio as
\begin{eqnarray}
    Y_B & \lesssim & 8.8 \times 10^{-28} \varepsilon 
    \left(\frac{\Omega_{mod}h^2}{0.2}\right)
    \left(\frac{m_{\phi}}{\mbox{TeV}}\right)^{-2/3}
    \left(\frac{m_{3/2}}{100\mbox{keV}}\right)^{2/3},
\end{eqnarray}
which is too small to explain the present value. 


\subsection{Unstable Q balls}

In the case of the unstable Q balls which decay into nucleons, it may
destroy light elements synthesized at the BBN. Thus, a new constraint
which the Q ball should decay before the BBN ($\sim 1$ sec), must be
imposed.


\subsubsection{Gauge-mediation type Q balls when the zero-temperature 
potential is dominated} 

There are several condition to be imposed. The first is the condition
that the Q ball is unstable, given by 
\begin{equation}
    \label{eq:instability}
    \frac{\phi_{osc}}{M} \lesssim 2.6 \times 10^{-6} 
    \left(\frac{M_F}{10^6\mbox{GeV}}\right)^2. 
\end{equation}
Second, the decay of the Q ball must be completed until the BBN,
otherwise it would spoil the success of the BBN, so that the life 
time $\tau_Q$ should be less than about 1~sec. Thus, we have
the following constraint:
\begin{equation}
    \label{eq:bbn}
     \frac{\phi_{osc}}{M} \lesssim 1.0 \times 10^{-6} 
     \left(\frac{M_F}{10^6 \mbox{GeV}}\right)^{6/5}.
\end{equation}
The largest possible value of $Y_B$ is
obtained by the conditions Eq.(\ref{eq:bbn}) and $M_F \lesssim
(m_{3/2}M)^{1/2}$, and will be written as
\begin{eqnarray}
    Y_B  & \lesssim & 2.8 \times 10^{-19} \varepsilon
    \left(\frac{\Omega_{mod}h^2}{0.2}\right)
    \left(\frac{m_{3/2}}{100\mbox{keV}}\right)^{4/5}.
\end{eqnarray}
This is thus too small to explain the present value $\sim 10^{-10}$.


\subsubsection{Gauge-mediation type Q balls when the thermal
logarithmic potential is dominated} 

Since the Q-ball charge is expressed as
$Q \simeq \beta (\phi_{osc}/M)^6 (T_{RH}/M)^{-4}$,
the unstable condition, $M_F Q^{-1/4} \gtrsim 1$ GeV, is given by
\begin{equation}
    \frac{\phi_{osc}}{M} \lesssim 3.4\times10^4
    \left(\frac{M_F}{10^6 \mbox{GeV}}\right)^{2/3}
    \left(\frac{T_{RH}}{M}\right)^{2/3},
\end{equation}
while the lifetime condition that the Q ball decays before the BBN
time ($\sim 1$ sec), is written as
\begin{equation}
    \label{eq:lifetime2}
     \frac{\phi_{osc}}{M} \lesssim 4.0\times10^3
     \left(\frac{M_F}{10^6 \mbox{GeV}}\right)^{2/15}
     \left(\frac{T_{RH}}{M}\right)^{2/3}.
\end{equation}

\begin{figure}[t!]
    \centering
    \includegraphics[height=7cm]{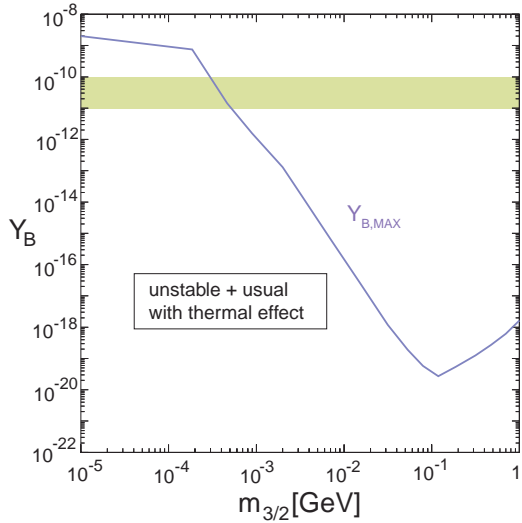}
    \caption[fig4]{\label{fig:unstusu} 
    Largest possible baryon-to-entropy ratio in the unstable Q-ball
    scenario in the thermal logarithmic potential.  }
\end{figure}

From the above conditions, we have the baryon-to-entropy
ratio as
\begin{equation}
    \label{eq:y-b-fig4}
    Y_B \lesssim 8.7 \times 10^{-11} \varepsilon 
    \left(\frac{\Omega_{mod}h^2}{0.2}\right)
    \left(\frac{m_{3/2}}{100\mbox{keV}}\right)^{-2/5}.
\end{equation}
We plot the
largest possible value of $Y_B$ [Eq.(\ref{eq:y-b-fig4})] in the
function of $m_{3/2}$ in Fig.~\ref{fig:unstusu}. As can be seen, we
can explain the present value in the range 
$m_{3/2}=10 - 100$ keV. However, the reheating temperature should be
as high as $5.6\times 10^{12}$ GeV, which may be rather higher for the
actual inflation models.

\subsubsection{Delayed Q balls}

Firstly, it is found that the unstable condition can be
expressed as $m_{3/2} \gtrsim 0.16$ GeV both when 
$V_T \lesssim V_{gauge}$ and when
$V_T \gtrsim V_{gauge}$.  Therefore, in
either case, we can estimate the baryon-to-entropy ratio as
\begin{equation}
    Y_B \lesssim 4.5 \times 10^{-17} \varepsilon
    \left(\frac{\Omega_{mod}h^2}{10^{-9}}\right)
    \left(\frac{m_{3/2}}{0.16\mbox{GeV}}\right)^{-1}
    \left(\frac{\phi_{osc}}{M}\right)^2,
\end{equation}
where $\Omega_{mod}h^2 \lesssim 10^{-9}$ for $m_{3/2} \gtrsim 0.16$
GeV from the X($\gamma$)-ray constraint. 
This is too small to explain the present 
value, even if $\phi_{osc} \sim M$.

\section{Conclusion}
We have investigate the possibility of the AD baryogenesis in the
gauge-mediated SUSY breaking scenario, while evading the cosmological
moduli problem by the late-time entropy production. In all the cases,
the Q-ball formation makes the efficiency of the baryon number
production considerably be diminish. In the zero-temperature potential 
$V_{gauge}$, whether the produced Q balls are stable or not, the
largest possible baryon-to-entropy ratio is too small to explain the
present value. This completely kills the successful situations
considered in Ref.~\cite{Asaka2}. 

We have also found that there are some marginally successful
situations when we take into account of the thermal effects on the
effective potential of the AD field. However, these successful
situations require very high reheating temperatures such as 
$10^{12} - 10^{16}$ GeV, which may be impossible to achieve in the
actual inflation models. 

In the delayed Q-ball formation case, we have found that enough
baryon-to-entropy ratio can be created, even in the zero-temperature
potential $V_{gauge}$ is dominant over the thermal logarithmic
potential $V_T$. It might be the unique solution for the AD
baryogenesis with solving the cosmological moduli problem, although
the scale of $M_F$ is rather low.

In addition, successful situations above need the following
conditions: the large initial amplitude such as $\phi_{osc} \simeq M$, 
and $\varepsilon \simeq 1$. This is realized if the A-terms, which
make the AD field rotate in the effective potential, originate from
some K\"{a}hler potential. Then, $\varepsilon \sim (\phi/M)^{\gamma} 
\sim 1$ for $\phi \sim M$, where $\gamma > 2$.

\Acknowledgements

This work is supported by the Grant-in-Aid for Scientific Research
from the Ministry of Education, Science, Sports, and Culture of Japan,
Priority Area ``Supersymmetry and Unified Theory of Elementary
Particles'' (No.\ 707).

\end{document}